# Large tensile strain induced monoclinic $M_B$ phase in BiFeO$_3$ epitaxial thin films on PrScO$_3$ substrate

Zuhuang Chen,[1,*] Yajun Qi,[1,2] Lu You,[1] Ping Yang,[3] C. W. Huang,[1] Junling Wang,[1] Thirumany Sritharan,[1] and Lang Chen[1,4,†]

[1] *School of Materials Science and Engineering, Nanyang Technological University, Singapore 639798, Singapore*

[2] *Key Laboratory of Green Preparation and Application for Materials, Ministry of Education, Department of Materials Science and Engineering, Hubei University, Wuhan 430062, P. R. China*

[3] *Singapore Synchrotron Light Source (SSLS), National University of Singapore, 5 Research Link, Singapore 117603, Singapore*

[4] *South University of Science and Technology of China, Shenzhen, China, 518055, P.R. China*

\* glory8508@gmail.com

† langchen@ntu.edu.sg

**Abstract**

Crystal, domain structures and ferroelectric properties of tensile-strained $BiFeO_3$ epitaxial films grown on orthorhombic $(110)_o$ $PrScO_3$ substrates were investigated. All films possess a $M_B$-type monoclinic structure with $109^o$ stripe domains oriented along the $[\bar{1}10]_o$ direction. For films thicknesses less than ~40 nm, presence of well-ordered domains is proved by the detection of satellite peaks in synchrotron x-ray diffraction studies. For thicker films, only the Bragg reflections from tilted domains were detected. This is attributed to the broader domain size distribution in thicker films. Using planar electrodes, the in-plane polarization of the $M_B$ phase is determined to be ~ 85 $\mu$C/cm$^2$，which is much larger than that of compressive strained $BiFeO_3$ films. Our results further reveal that the substrate monoclinic distortion plays a major role in determining the stripe domain formation of the rhombohedral ferroic epitaxial thin films, which sheds light to the understanding of elastic domain structure evolution in many other functional oxide thin films as well.

## I. INTRODUCTION

Epitaxial strain is recognized as an effective parameter to modify the structure and tune many physical properties of functional oxide thin films.[1] Among the functional oxides, BiFeO$_3$ (BFO) provides good opportunities for tunable functionalities through its rich strain-temperature related phase diagram because of the inherent coupling among elastic, electric, and magnetic order parameters.[2] For this reason, the effect of strain in BFO epitaxial thin films has received considerable attention, leading to several interesting theoretical predictions and experimental verifications in strained BFO films.[3-9] At room temperature, bulk BFO possesses a rhombohedrally distorted perovskite structure with space group $R3c$ having pseudocubic lattice parameters, $a = 3.965$ Å and $\alpha = 89.4°$.[10] (Pseudo-cubic index is used throughout this letter unless otherwise specified) When BFO film is epitaxially grown on low-mismatch single crystal substrate with equal in-plane lattice parameters, its symmetry is lowered to monoclinic due to the in-plane biaxial strain.[11,12] There are three different types of monoclinic ferroelectric phases, namely, $M_A$, $M_B$ and $M_C$, following the notation of Vanderbilt and Cohen.[13] Among them, the $M_A$ and $M_B$ phases belong to the space group $Cm$ (or $Cc$), and the $M_C$ phase belongs to $Pm$ (or $Pc$). The differences between the $M_A$ and $M_B$ phases lie in the magnitudes of their monoclinic lattice parameters and consequently their polarization components corresponding to the pseudocubic unit cell: for the $M_A$, $a_m/c_m < \sqrt{2}$, and $P_x = P_y < P_z$; whereas for the $M_B$, $a_m/c_m > \sqrt{2}$, and $P_x = P_y > P_z$. (Subscript $m$ denotes monoclinic indices) The orientation of polarization vector is, respectively, along $[uuv]$ ($u < v$), [$uuv$] ($u > v$), and [$u0v$] in the monoclinic $M_A$, $M_B$, and $M_C$ phases. As shown in Figure 1, these low-symmetry monoclinic phases could act as structural bridges linking the tetragonal (*T*), rhombohedral (*R*), and orthorhombic (*O*) phases, and the monoclinic symmetry allows the polarization vector to be unconstrained within certain symmetric plane.[14] Therefore, the polarization in monoclinic phase could continuously rotate in the symmetric plane under an external stimulus, such as electric field,[15] pressure[16] or epitaxial strain[4,17]. Such symmetry-allowed polarization rotation has been proposed to be responsible for the high piezoelectric response of ferroelectric monoclinic phases.[18] Recent experiments have demonstrated a phase sequence of $R \rightarrow M_A \rightarrow M_C \rightarrow T$ with increasing compressive strain in (001)-oriented BFO epitaxial thin film.[7,19] And the compressive-strain-stabilized, highly distorted tetragonal-like $M_C$ phase could coexist with rhombohedral-like $M_A$ phase, forming a morphotropic phase boundary with the potential of enhanced ferroelectric and piezoelectric responses.[5]

In comparison to the extensive reports on the effect of compressive strain in BFO epitaxial films, the tensile strain effect is less investigated. One reason is that most commercially available perovskite single crystal substrates have in-plane lattice parameters smaller than that of bulk BFO. Recent phenomenological calculations predicted that a tensile strain larger than ~ 1.5% could induce a new, orthorhombic *O* phase whose polarization vector lies along the in-plane [110] direction.[5,20-22] In contrast, more recent first-principles calculations reported that the orthorhombic phase transition only occurs at a tensile strain larger than 5%.[23,24] However, the rhombohedral $R$ phase might not transform directly to orthorhombic $O$ phase as there is a bridging monoclinic $M_B$ phase between $R$ and $O$. Experimental verification of the existence of the $M_B$ phase and tensile-strain induced phase transition/polarization rotation could now be done using the new, rare-earth scandate $PrScO_3$ (PSO) single crystal substrate which has an orthorhombic structure with lattice constants $a_o = 5.780$ Å, $b_o = 5.608$ Å, and $c_o = 8.025$ Å.[25] (Subscript $o$ denotes orthorhombic indices) The orthorhombic unit cell can be related to a tilted pseudo-cubic (monoclinic) unit cell through the following relationships:

$$a = \frac{c_o}{2} = 4.013 \text{ Å}, b = c = \frac{\sqrt{a_o^2 + b_o^2}}{2} = 4.027 \text{ Å}, \alpha = 2\arctan\frac{b_o}{a_o} = 88.3°, \beta = \gamma = 90°$$

It is important to note that the $(110)_o$-oriented PSO substrate has the largest in-plane pseudo-cubic lattice parameters among all commercially available perovskite single crystal substrates. PSO could impose tensile strains of 1.2% and 1.6% along the $[001]_o$ and $[\bar{1}10]_o$ directions, respectively, in BFO films. Therefore, PSO was chosen as the substrate for investigating the effect of tensile strain in BFO films in this work.

Here, we report a detailed structure characterization of the tensile-strained BFO films with various thicknesses grown on PSO single crystal substrates, and determine the role of substrate monoclinic distortion on the domain formation in the tensile strained BFO films. We also demonstrate that the tensile strain induces a rotation of the spontaneous polarization toward in-plane direction by directly measuring the in-plane polarization of the films.

## II. EXPERIMENTAL AND COMPUTATIONAL METHODS

BFO films with thicknesses ranging from 8 to 110 nm were deposited by pulsed laser deposition (PLD) on (110)$_o$-oriented PSO substrates. The deposition temperature and the oxygen pressure were 700 °C and 100 mTorr, respectively.[7,26] High resolution x-ray diffraction (HR-XRD) measurements were performed at the Singapore Synchrotron Light Source (SSLS). The diffraction data was plotted in reciprocal lattice units (r.l.u.) of the PSO substrate defined by the monoclinic unit cell (1 r.l.u. = $2\pi/4.013$Å$^{-1}$ for *H* direction, 1 r.l.u. = $2\pi/4.027$Å$^{-1}$ for *K* and *L* directions). *H*, *K*, and *L* are reciprocal space coordinates. The thickness of the BFO film was calibrated by x-ray reflectivity (XRR).[27] The surface morphology and piezoelectric force microscopy (PFM) investigations were carried out on an Asylum Research MFP-3D atomic force microscope (AFM). In-plane ferroelectric hysteresis loops were measured at a frequency of 1 kHz using a Precision LC ferroelectric tester (Radiant Technologies) on planar electrode devices, where Pt electrodes were patterned on top of the BFO films.[17,27] Transmission electron microscopy (TEM) was done in a JEOL 2100F microscope at 200 kV.

## III. RESULTS AND DISCUSSION

Figure 2(a) displays a typical *L*-scan (specular $\theta-2\theta$ scan) of a 20-nm-thick BFO film. Only the 00*L* peaks of the PSO substrate and the BFO film are detected, indicating phase-pure and epitaxial growth. The thickness fringes apparent near the BFO diffraction peaks indicate the high quality of the films. The out-of-plane *c* lattice parameter calculated from the 00*L* peaks is ~ 3.89 Å, which is smaller than the bulk value of BFO due to the Poisson's effect arising from the in-plane tensile strain. The *c* lattice parameter determined from the 00*L* Bragg reflections for all film thicknesses did not vary signifying that the tensile strain is retained even in the thicker films. Figure 2(b) is a typical AFM topographic image of a 50 nm film showing a smooth film surface. Both out-of-plane [Fig. 2(c)] and in-plane PFM [Fig. 2(d)] images, taken with the cantilever along [110], show a stripe-like contrast over large areas. Such stripe domain patterns with domain walls along $[\bar{1}10]_o$ direction, and the corrugated film surface are typical features of 109° periodic domain configuration in rhombohedral BFO films (note that we still use notation of 109° domains in bulk BFO although the real angle between polarizations of neighboring domains in epitaxial films is strain-dependent).[28-32]

It is known that PFM technique is hard to provide sufficient resolution to detect domain structure in ultrathin ferroelectric films because of the weak piezoelectric response and fine domain feature of the ultrathin films. On the other hand, synchrotron x-ray scattering is a powerful technique to probe the crystal and domain structures of ferroic ultrathin films.[33-36] The different cation shift within each domain gives rise to different unit cell structure factors. If the domains are well ordered, the difference in the structure factors between adjacent domains leads to satellite peaks at specific reciprocal lattice vectors which have a component parallel to the relative polar shift direction.[33-39] Therefore, analysis of the distribution and orientation of the satellite peaks by HR-XRD could provide additional information regarding the domain wall orientation and polar symmetry in ferroelectric thin films.[33,34] We embark on such an exercise here.

Figure 3 shows the specular and off-specular reciprocal space mappings (RSMs) for a 13-nm film obtained by synchrotron HR-XRD. As evident in Figs. 3(a) and 3(b), only two Bragg spots with identical *K* values from the substrate and the film are detected in the (0*KL*) zone when the incident x-ray beam is along the $[\bar{1}10]_o$ direction, indicating that there is no tilt between the (001) planes of the film and the substrate along the $[\bar{1}10]_o$ direction and the film is coherently strained. When the incident x-ray is aligned parallel to the $(001)_o$ direction, satellite peaks appear with equal spacing in reciprocal space in both (002) and ($\bar{1}$03) mappings as clear in Figures 3(c) and 3(d). This implies that the satellite peaks are not from Bragg reflections of tilted domains but, are from periodic domain modulation. The presence of satellite peaks in the (00*L*) and (0*K*0) (see supplementary materials, ref. 27) zones indicates that spontaneous polarization in the tensile-strained film has both out-of-plane as well as in-plane components. This eliminates the possibility of the structure belonging to the orthorhombic *aa* phase with purely in-plane polarization.[33,34] The possible domain structure, that is 109º stripe domains with walls along $[\bar{1}10]_o$ direction, are illustrated in Fig. 4(a). Furthermore, it is found that the BFO (0$\bar{1}$3) reflection has a larger *L* value than the BFO (003) and (013) counterparts (diffraction data not shown here). Then, the *α* angle of the BFO pseudocubic unit cell must deviate from 90°, and the pseudocubic unit cell is tilted along the $[\bar{1}10]_o$ direction by an angle of $\delta = 90 - \alpha_{BFO} = \arctan\frac{L_{0\bar{1}3} - L_{013}}{2} \sim 1.6°$. That is, the *α* angle of the BFO unit cell follows the monoclinic distortion of the substrate, as shown in Fig. 4(b). Previous phenomenological modeling showed that all eight polarization variants in rhombohedral BFO remain degenerate even under an large substrate-induced in-plane anisotropic strain.[40] Therefore, we suggest that the domain degeneracy is broken by the monoclinic distortion propagated from the substrate, rather than by the in-plane anisotropic strain, leading to two-variant stripe domains with net shear direction along the substrate monoclinic distortion direction.

Figure 5 shows RSMs for a 93-nm-thick BFO film. The diffraction data of the film shows no peak splitting in (0$KL$) zone but splitting occurs in ($H$0$L$) zone, a feature similar to that reported in BFO films with 109° domains grown on other rare-earth scandate substrates with relatively smaller lattice parameters.[29,30,32] Peak splitting occurring only in the ($H$0$L$) zone can be attributed to the generation of two Bragg diffraction peaks from two tilted structural variants, where the (00$L$) plane is tilted by an angle of ~±0.4° in [001]$_o$ with respect to the substrate surface. Combining all diffraction information one could show that a monoclinic phase of BFO with derived lattice parameters of $a_m$ = 5.741(7) Å, $b_m$ = 5.634(7) Å, $c_m$ = 3.884(3) Å, $\beta_m$ = 89.29(7)°, could explain all the observed structural information (subscript $m$ denotes monoclinic indices).[27] The fact $c_m < b_m/\sqrt{2} < a_m/\sqrt{2}$ indicates that the monoclinic phase is a $M_B$-type structure. The polarization in the $M_B$ phase lies within the ($1\bar{1}0$) plane and tilts away from the [111] direction, towards the in-plane [110] direction.[7] To attain the theoretically-predicted, *orthorhombic aa* phase, further tilt of the polarization vector down towards the in-plane direction is necessary. Therefore, a larger tensile strain is required to achieve such tilting of the polarization vector and stabilize the orthorhombic phase.[23,24]

In-plane *H* scans (rocking curves) around the 00*L* BFO reflections were done to understand the origin of the diffraction intensity distribution and structural evolution with film thickness. Figure 6 displays the *H*-scan curves for different diffraction orders of films. As clear in Figure 6a, the 13-nm-thick film exhibits satellite peaks with equal spacing up to the second order in all reflections, indicating that the domains are well-ordered with a very narrow size distribution. It is important to note that the central zero-order peak with $H = 0$ originates from the modulation structure, i.e., $0^{th}$ order of the satellite peak.[35] The intensity of Bragg peak is hidden by the satellite peaks. At intermediate film thickness of ~35 nm, satellite peaks and Bragg peaks are simultaneously observed as evident in Figure 6b. Only the first order satellite peaks are visible in this figure, indicating that the domains are less ordered and thus have a larger size distribution than those in the 13 nm thick film. The spacing Δ*H* between the satellite peaks yields the domain modulation periodicity *D* in the real space. And it is found that *D* increases with increasing film thickness for films thicker than ~10 nm, for instance, *D*=0.4013 nm*/*Δ*H* ≈ 20 and ≈ 45 nm for the 13 and 36 nm BFO film, respectively. With further increasing film thickness, as shown in Figure 5c, for a 93-nm-thick film, the spacing between the two tilted peaks increases with the diffraction order *L*, suggesting that two distinct Bragg diffraction peaks originate from the tilted domains. Absence of clear satellite peaks in the thick films indicates that the domain size distribution in these films is wide and scattered.

The domain structures of the films were further examined using TEM. Figure 7(a) is a typical plan-view TEM image of an 8-nm-thick film. A stripe-like domain pattern with walls along the $[\bar{1}10]_o$ direction clear in the TEM image of the 8 nm film was also observed in thicker films by PFM. Figure 7(b) shows cross-sectional dark field TEM images of a 20 nm thick film. Domains with alternating dark and bright contrasts separated by vertical $(001)_o$ domain walls are observed proving the presence of 109° domain pattern in the film. The domain widths ranged from 6 nm to 24 nm and the average size of the domains is between 11 and 14 nm, which is close to the half of modulation periodicity determined from the HR-XRD data. Figures 7(c) and 7(d) show two typical selected-area electron diffraction (SAED) patterns taken from the interface area and the top area of the film viewed along $[\bar{1}10]_o$ or $[010]_c$ zone axis. From the pattern, the orientation relationship of the film and the substrate can be determined to be $(001)_c$ BFO//$(110)_o$ PSO and $[010]_c$ BFO//$[110]_o$ PSO. The diffraction pattern in Figure 7(d) shows the splitting on high order reflection spots, indicated by arrows. Such splitting is consistent with the existence of two ferroelastic variants ($r_1$ and $r_4$) in the films revealed by PFM and XRD studies.

We have studied in-plane domain switching and also directly measured in-plane polarization hysteresis loops of compressively strained BFO films using planar electrode devices and remnant hysteresis measurements.[17,41] Here, in-plane polarization of the tensile-strained films is determined using the same method. Figure 8 shows in-plane P-E loops of 50 nm film grown on PSO. The loops are well saturated and rectangular in shape. It is clear that the in-plane polarization value and coercive field of the $M_B$ phase obtained in the film on PSO are much larger than those of the compressive strained film on SrTiO$_3$. The remnant polarization ($P_{E//[100]}$) of the $M_B$ phase obtained in the film on PSO is ~ 60 $\mu C/cm^2$ for $E$ lying along [100], which is in good agreement with the polarization value calculated by first-principle calcuations for the film under the same tensile strain.[24] Considering that the spontaneous polarization in monoclinic $M_B$ phase is constrained within the ($1\bar{1}0$) plane, therefore, the in-plane polarization in each individual domain is along <110> and the net in-plane polarization of stripe domains is the sum of their resolved components in <100>. Thus, the in-plane polarization ($P_{in}$) of each domain in the $M_B$ phase is then

$P_{in} = \sqrt{2} P_{E//[100]}$ ~ 85 $\mu C/cm^2$, which is much larger than that in compressive strained films.[17] Such a large in-plane polarization in the tensile-strained film is probably due to the tensile strain induced a rotation of spontaneous polarization toward in-plane [110] direction, and thus has a larger in-plane projection.

## IV. CONCLUSION

In summary, we have determined that tensile-strained BFO films grown on $(110)_o$ PrScO$_3$ single crystal substrates have the unit cell symmetry of monoclinic $M_B$ phase and consist of stripe domains separated by 109 ° domain walls. Films with thickness less than 40 nm contain well-ordered nano-domains which give distinct satellite peaks in synchrotron HR-XRD. Satellite peaks are not clearly observed in thicker films because the domain sizes are larger and their size distributions are broader. We also show that the substrate monoclinic angle plays a major role in determining the stripe domain formation of rhombohedral BFO thin films. Furthermore, by directly measured the in-plane polarization of the $M_B$ phase, we found that the large tensile strain induces the rotation of the spontaneous polarization toward in-plane direction and gives rise to a large in-plane polarization component of ~ 85 $\mu$C/cm$^2$.

**Acknowledgement**

We acknowledge the supports from the Singapore National Research Foundation under the Campus for Research Excellence And Technological Enterprise (CREATE) programme: Nanomaterials for Energy and Water Management. Z.C. acknowledges the support from Ian Ferguson Postgraduate Fellowship and thanks A. R. Damodaran for some useful discussions. L.C. also acknowledges support from the 9$^{th}$ China Thousand Talent program and a start-up fund from SUSTC. Y.Q. acknowledges the support from the National Science Foundation of China (11204069). Part support is from SSLS via NUS Core Support C-380-003-003-001.

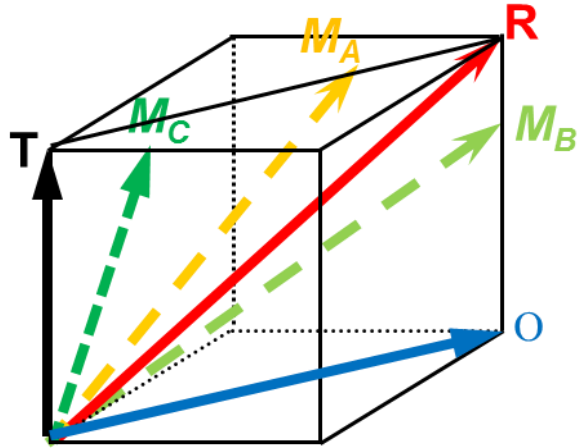

FIG. 1. (Color online) Sketch of the relationship between T, O, R, and three M phases. T, R, O, and M represent tetragonal, rhombohedral, orthorhombic and monoclinic phases, respectively. Arrows represent the directions of the spontaneous polarization *P*.

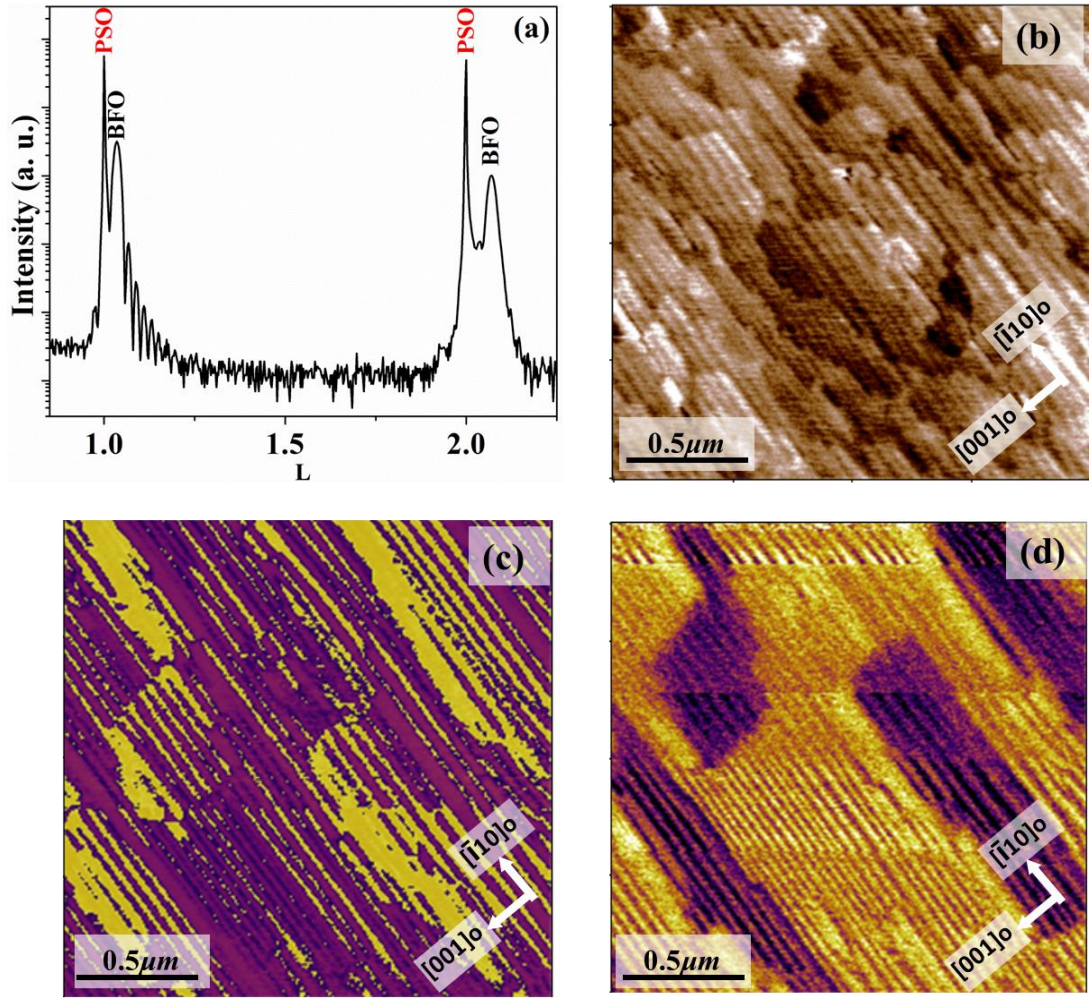

FIG. 2. (Color online) (a) *L* scan along (00*l*) for a 20 nm BFO film grown on PSO. (b) A typical AFM topographic image of the film. (c) Out-of-plane, and (d) In-plane PFM images of the BFO film.

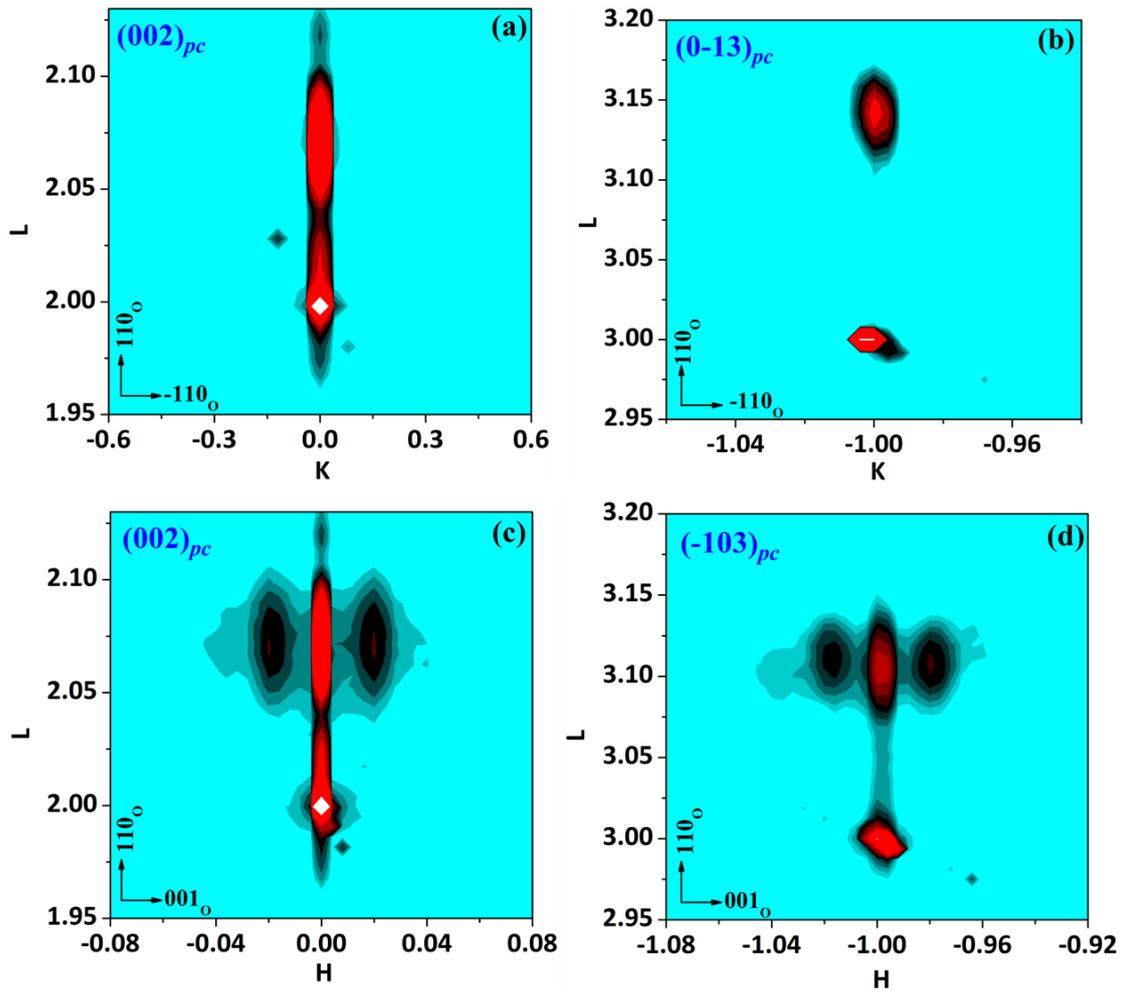

FIG. 3. (Color online) *KL* reciprocal space mappings around (a) (002) and (b) (0$\bar{1}$3); and *HL* reciprocal space mappings around (c) (002) and (d) ($\bar{1}$03) of a 13-nm-thick BFO film.

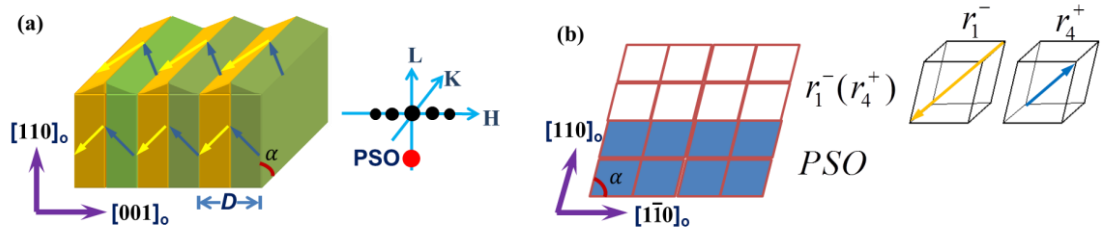

FIG. 4. (Color online) (a) (left) Schematic of 109°-like periodic stripe domains, and (right) the (00$L$) associated peak positions of the film (black circles) and the substrate (red circle) in reciprocal space. (b) Side view along [001]$_o$ of the twined domain.

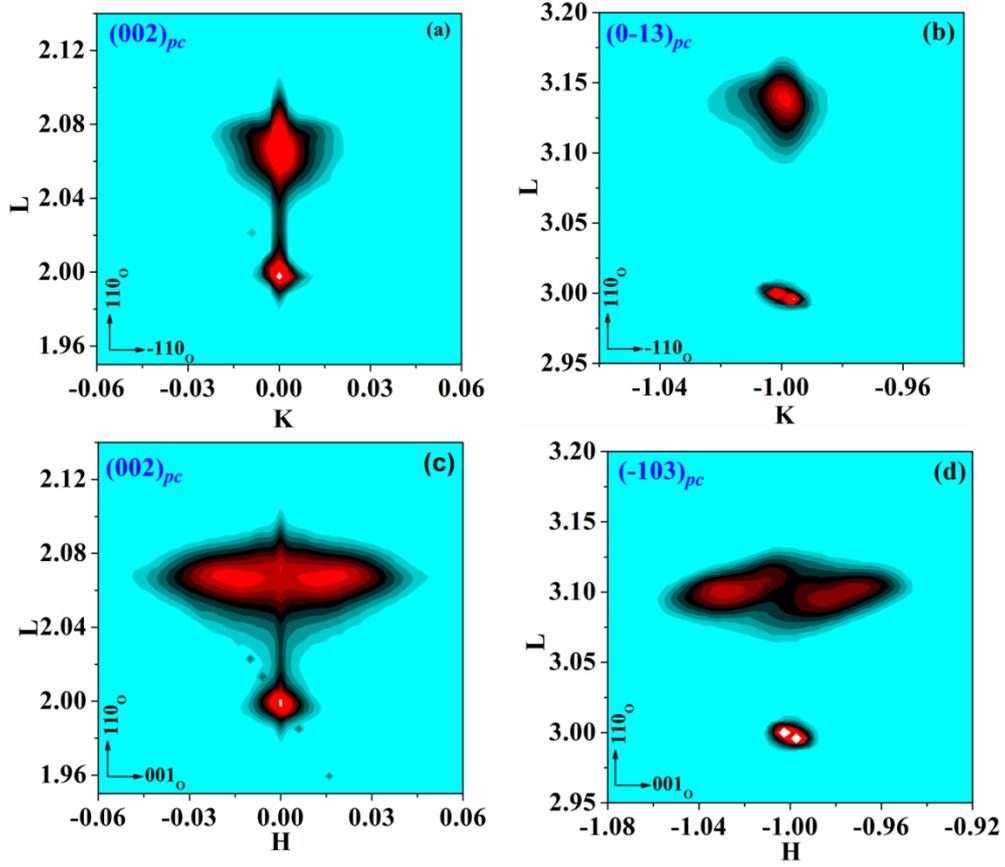

FIG. 5. (Color online) *KL* reciprocal space mappings around (a) (002) and (b) (0$\bar{1}$3); and *HL* reciprocal space mappings around (c) (002) and (d) ($\bar{1}$03) of a 93-nm-thick BFO film.

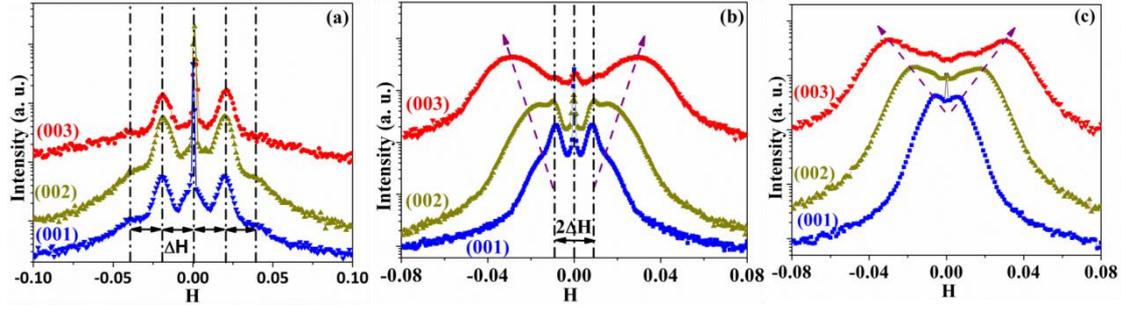

FIG. 6. (Color online) *H* scans around BFO (00*L*) for (a) 13 nm, (b) 36 nm, and (c) 93 nm films. (Dash lines are guides to the eyes)

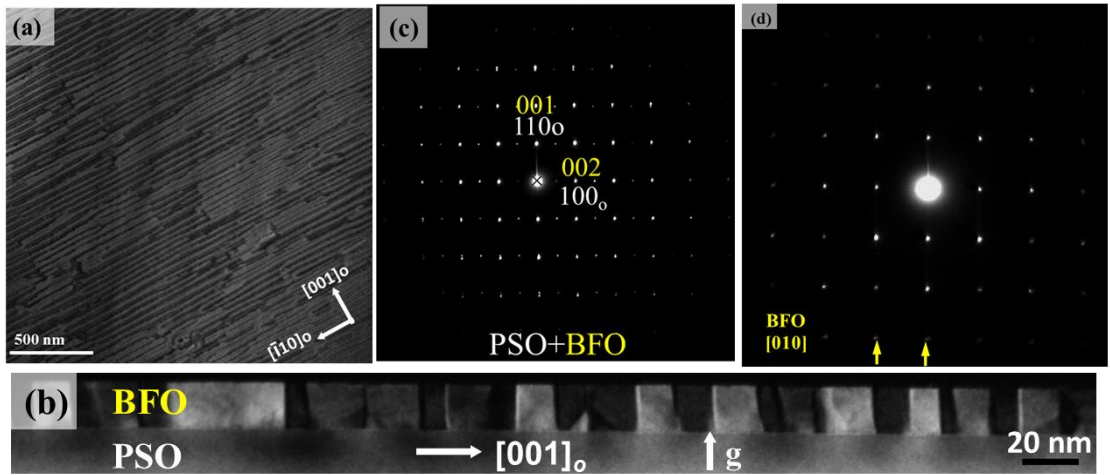

FIG. 7. (Color online) (a) Typical plan view bright-field TEM image of an 8 nm thick BFO film. (b) Cross-sectional dark-field TEM images of a 20 nm thick film. (c) SAED pattern taken from the interface between the BFO film and the PSO substrate. (d) SAED pattern taken from the top area of the BFO film, the splitting on high-order reflection spots is indicated by arrows.

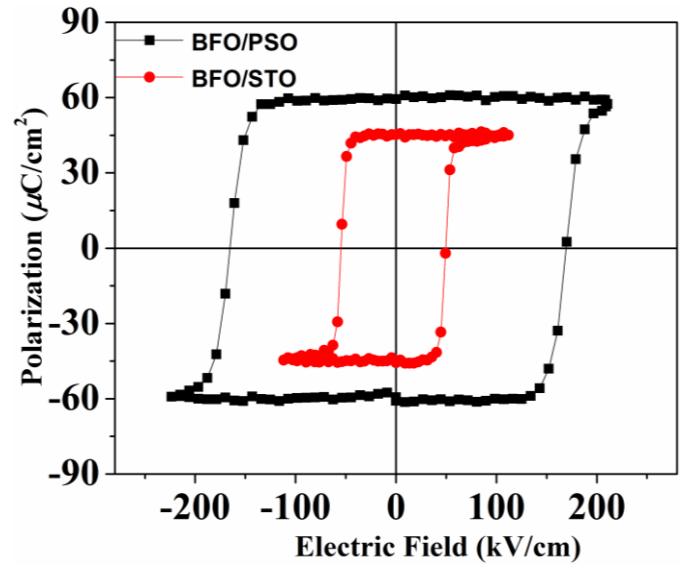

Figure 8: Field-dependent variation of in-plane *P-E* loops of 50 nm BFO films grown on PSO and SrTiO$_3$ substrate when the electric field is applied along the [100] direction.

# Supplementary Material

# Monoclinic $M_b$ phase in tensile-strained BiFeO$_3$ epitaxial thin films on PrScO$_3$ substrate

Zuhuang Chen,[1] Yajun Qi,[1,2] Lu You,[1] Ping Yang,[3] C. W. Huang,[1] Junling Wang,[1] Thirumany Sritharan,[1] and Lang Chen[1,4]

[1] *School of Materials Science and Engineering, Nanyang Technological University, Singapore 639798, Singapore*

[2] *Key Laboratory of Green Preparation and Application for Materials, Ministry of Education, Department of Materials Science and Engineering, Hubei University, Wuhan 430062, P. R. China*

[3] *Singapore Synchrotron Light Source (SSLS), National University of Singapore, 5 Research Link, Singapore 117603, Singapore*

[4] *South University of Science and Technology of China, Shenzhen, China, 518055, P.R. China*

## S1. Determining film thickness by synchrotron X-ray reflectivity (XRR)

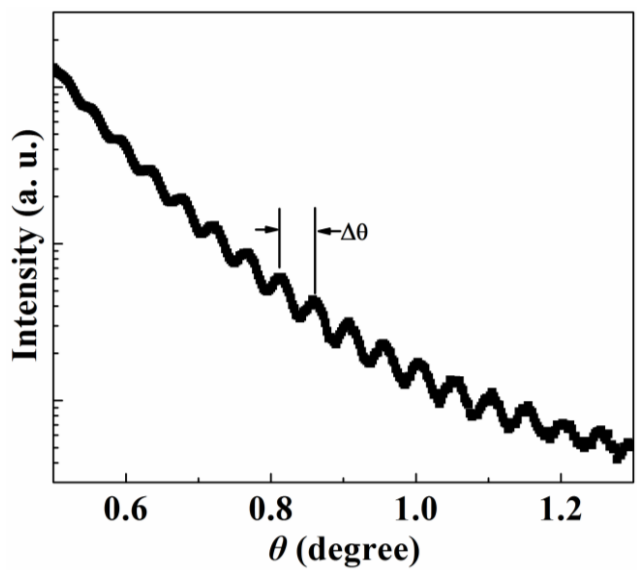

**Figure S1: A typical XRR curve of a BFO film grown on PrScO$_3$ substrate.**

A typical XRR curve of a BFO film grown on PSO substrate is shown in Figure S1. The presence of Kiessig fringes indicates a high quality sample with a smooth surface. The film thickness $d$ is calculated to be 93 nm by the period of the intensity oscillations.

## S2. Grazing incidence synchrotron X-ray diffraction

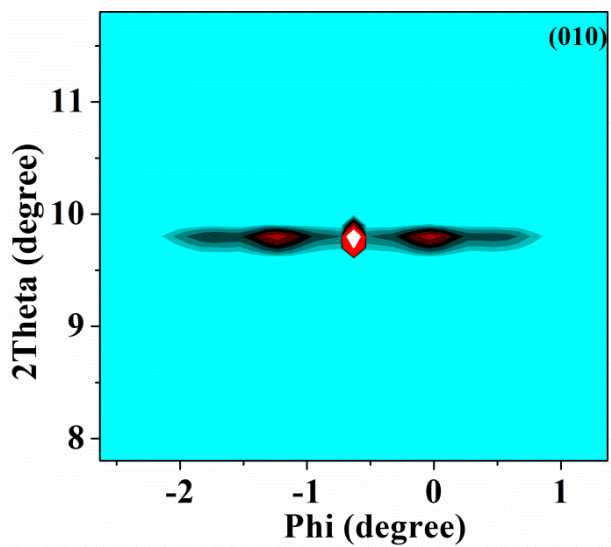

**Figure S2: In-plane reciprocal space map around the (010) reflection of the 13 nm BFO film grown on PSO.**

In order to get more information about the polar symmetry of the strained BFO films, grazing incidence X-ray diffraction (GIXRD) was performed at Shanghai Synchrotron Radiation Facility (SSRF) to map out the in-plane reciprocal space. Figure S2 shows reciprocal space map around the 010 Bragg peak of the 13 nm BFO film grown on PSO. Satellite peaks are observed around 010 peak, indicating that the polarization has an in-plane component.

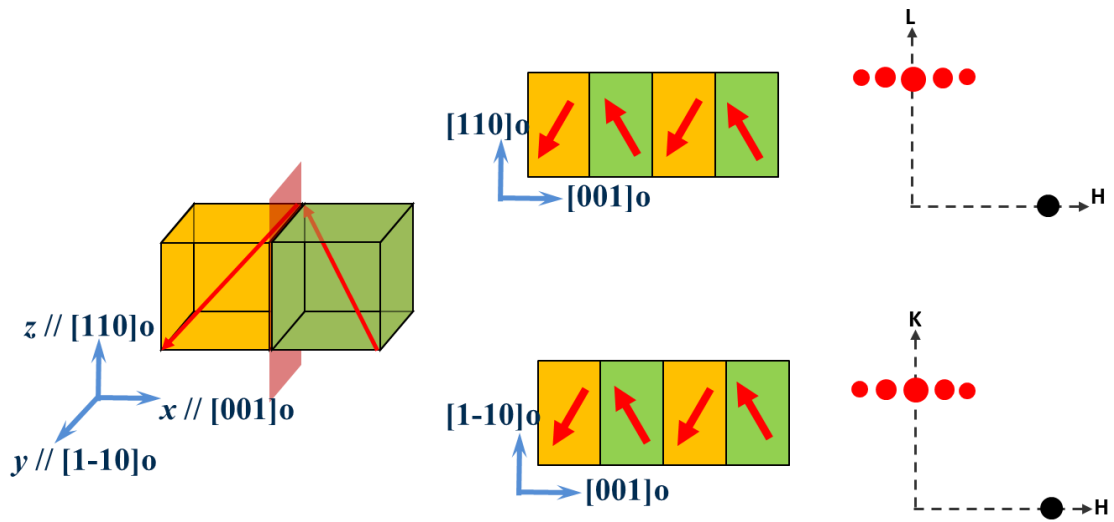

**Figure S3: Schematic representation of 109° stripe domains and the observed reciprocal space mappings.**

Since satellite peaks exist around (00L), (H0L) and (0K0) reflections in the H direction, spontaneous polarization in the tensile strained films must have both out-of-plane and in-plane components. Therefore, although the Bragg peaks are hidden under the satellite peaks, we propose that the ultrathin BFO film grown on PSO is monoclinic $M_B$ phase (spontaneous polarization component $P_x = P_y > P_z$, if we neglect the in-plane anisotropic strain) and has the same crystal structure as the thick film. Figure S3 shows schematically how periodic 109° stripe domain can give rise to satellite peaks. As there is no difference in spontaneous polarization vector in [001]$_o$ direction, i.e., there is no difference in $P_x$, no modulation is visible (extinct) around the (H00) reciprocal spots, as shown in Figure S4.

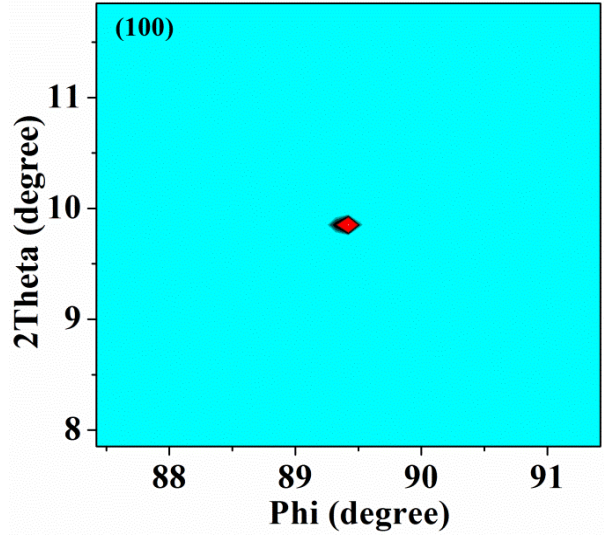

**Figure S4: In-plane reciprocal space map around the (100) reflection of the 13 nm BFO film grown on PSO.**

**S3. Determination of lattice parameters of the 93 nm thick film**

We can determine the lattice parameters and crystal system of the thick film by measuring the coordinates for several reciprocal space vectors (RSVs) [For instance, (002), (0$\bar{1}$3), (013), and ($\bar{1}$03)]. Mathematical operation of these RSVs will lead to reduction of three basis vectors, $a^*$, $b^*$, and $c^*$ in the reciprocal space. The lengths and angles of $a$, $b$, and $c$ can subsequently be obtained in the real space and hence the crystal system is concluded. The coordinate data of the PSO substrate and the BFO film are listed in Table 4-1.

Combing all the RSMs, particularly using RSVs, we can determine the lattice parameters of the film. The basis vectors (crystallographic axes) were firstly deduced from precisely measured coordinates for RSVs (002), (0$\bar{1}$3) and ($\bar{1}$03) as discussed above. The lattice parameters for the pseudo-cubic unit cell of the 93-nm-thick film are determined to be: $a_{pc}$ = 4.008 Å, $b_{pc}$ = 4.027 Å, $c_m$ = 3.885 Å, $\alpha$ = 89.62 °, $\beta$ = 90.47 °, and $\gamma$ = 91.20 °, leading to a monoclinic unit cell with $a_m$ = 5.741(7) Å, $b_m$ = 5.634(7) Å, $c_m$ = 3.884(3) Å, $\beta_m$ = 89.29 (7) °.

**Table S1: Coordinate data in reciprocal space of the PSO substrate and the 93-nm-thick film.**

|       | PSO | BFO |
|-------|-----|-----|
| (002) | (0.0016, 0.0000, 1.9998) | (0.0140, 0.0000, 2.0724) |
| ($\bar{1}$03) | (-1.0000, 0.0000, 3.0019) | (-0.9807, 0.0000, 3.1073) |
| (0$\bar{1}$3) | (0.0000, -1.0000, 2.9985) | (0.0000, -1.0001, 3.1454) |

**S4. Domain size distribution of films with different thickness**

The increase of domain size distribution with increasing film thickness was further directly demonstrated by TEM. Figure S5 shows cross-sectional dark field TEM images of two films of thicknesses (a) 20 nm and (b) 110 nm, respectively. The distribution of the domain sizes measured from over 15 TEM images is shown in the Fig. S5(c) which was taken from thin specimens of ~8 $\mu$m length of the films. For the 20 nm film, the domain widths ranged from 6 nm to 24 nm and the average size of the 109° domains is between 11 and 14 nm, which is close to the half of modulation periodicity determined from the XRD data. While for the 110 nm film, the domain width distribution was relatively broader ranging from 15 to 91 nm.

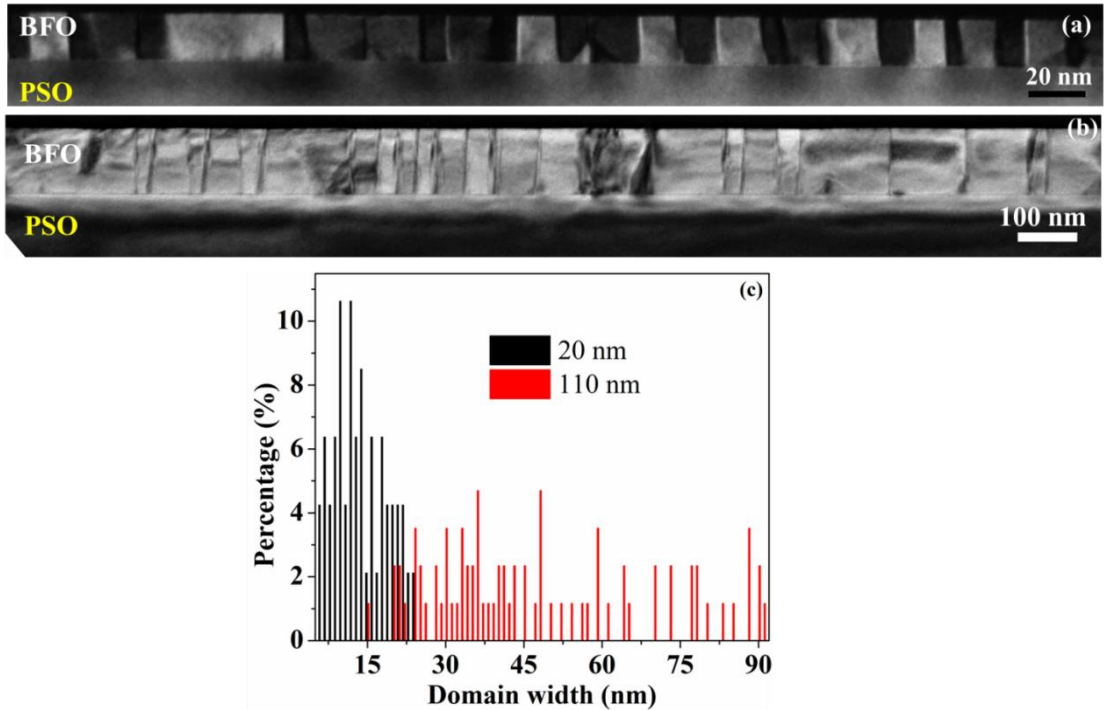

**FIG. S5.** Cross-sectional dark-field TEM images of (a) 20 nm and (b) 110 nm films. (c) Statistical domain size distribution of the two films.

**S5. Temperature dependence of domain satellites**

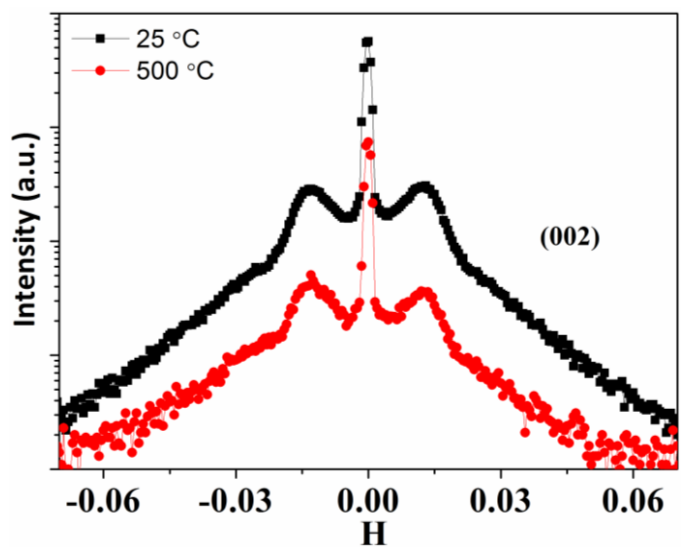

**FIG. S6.** *H* scans around the 002 diffraction peak of BFO at 25 °C and 500 °C of a 23-nm-thick BFO film on PSO.

Figure S6 shows rocking curve $H$ scans around the BFO (002) reflection of a 23-nm-thick film grown on PSO substrate at room temperature and at 500 $^o$C which is the highest temperature that we can achieve. It can be seen that the satellites due to periodic twinning are still visible at 500 $^o$C and the satellite spacing is almost the same as that at room temperature, which indicates that the Curie temperature of the $M_B$ phase could be above 500 $^o$C at least.

**S6. In-plane polarization and domain switching of the tensile strained BFO films**

In order to determine the polarization value in the $M_B$ phase obtained in tensile-strained films, and also to demonstrate on the effect of strain on polarization, we directly measured the in-plane *P-E* loops of BFO films using remnant hysteresis measurement. Planar Pt electrodes were patterned on BFO films via a standard lift-off procedure. The edge of the Pt electrode was aligned along the [001]$_o$ (or [100]) and [$\bar{1}$10]$_o$ (or [010]) of the substrate. The norminal channel width is 5 um.

For planar capacitor with electrodes parallel to 109 ° domain walls, which is along [$\bar{1}$10]$_o$, only a linear P-E loop was seen and no domain switching occurs after applied a maximum voltage of 99 V. For planar capacitor with electrodes perpendicular to 109 ° domain walls, a square P-E loop was seen (see Figure 8 in the main content) and 109 ° domain pattern actually turns into 71 ° domain pattern after applying an electric field of 99 V, as shown in Fig. S7.

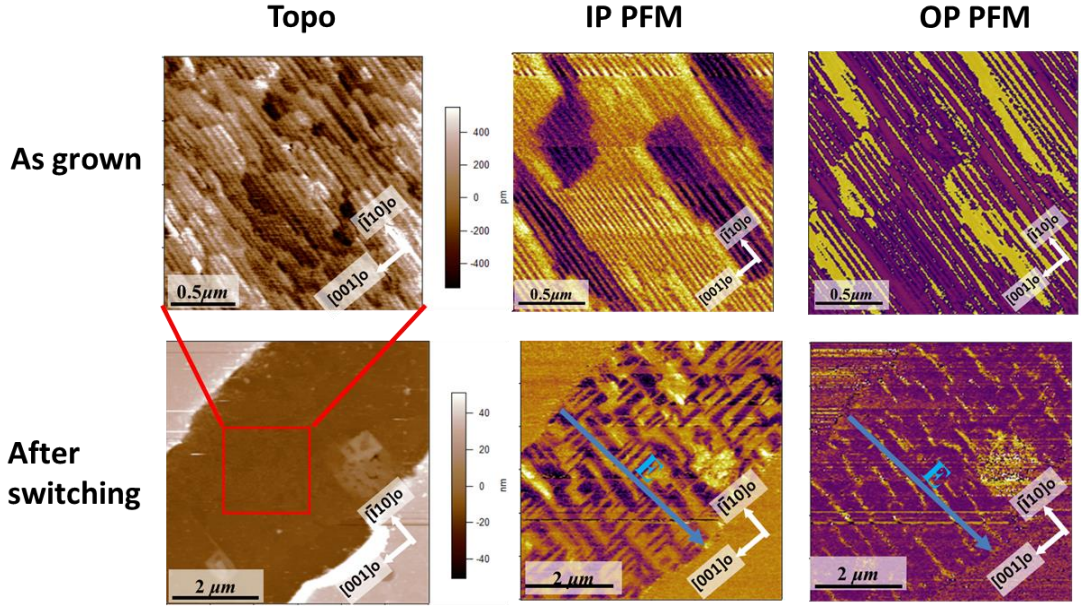

**FIG. S7. Topography, In plane PFM, and out of plane PFM images of quasi-planar BFO capacitors fabricated on films grown on PSO substrates with electrode edges along $[001]_o$ directions.**

**S7. x-ray diffraction of BFO thin films grown on different substrates**

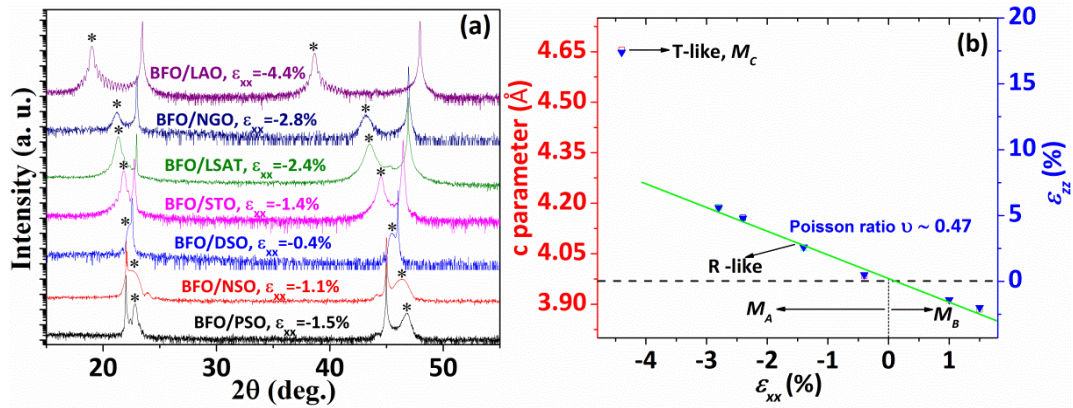

**Figure S8: (a) $\theta$-$2\theta$ scans of (001)-oriented ~ 30-nm-thick BFO films grown on different substrates. The * denotes the peak position of BFO. (b) The out-of-plane $c$ lattice parameters (red) and lattice strains $\varepsilon_{zz}$ ( blue) plotted as a function of the in-plane misfit strain $\varepsilon_{xx}$.**

Figure S8(a) shows HRXRD $\theta-2\theta$ scans of ~30-nm-thick (001)-oriented BFO films grown on different substrates. The averaging in-plane lattice parameters of the films determined by off-axis RSMs are close to those of the underlying substrates, which suggest that the films are fully strained in spite of a large variation of the substrate lattice constants. The sudden shift in the peak position between films on NGO and films on LAO evident is attributed to a first-order structural phase transition (R-like → T-like). Fig. S8(b) shows the $c$ lattice parameter plotted versus the in-plane misfit strain $\varepsilon_{xx}$ and out-of-plane lattice strain $\varepsilon_{zz}$. Note that the data points for all the films except on LAO show a linear relationship between the $c$-parameter (or $\varepsilon_{zz}$) and $\varepsilon_{xx}$, which indicates that the misfit strains can be retained without relaxation and a pure elastic deformation. If a first order R-like monoclinic → orthorhombic phase change occurs in the tensile strained films, we should observe a sudden jump in the c lattice parameter (or $\varepsilon_{zz}$) rather than a linear evolution as shown in the Figure S8b. Therefore, we could infer that, for BFO films under misfit strains of ~+1.5% (on PSO substrate), the R-like monoclinic $M_B$ phase is obtained.

Poisson ratio $v$ of the R-like phases can be calculated using the equation $v = 1/(1 - 2\varepsilon_{xx}/\varepsilon_{zz})$. $\varepsilon_{xx}/\varepsilon_{zz}$ is estimated by a linear fit (green line) to the data points in Figure S8b which gives the value of $v = 0.47$. This is the largest range of epitaxial strain over which $v$ has ever been determined for BFO obtained in high quality coherent-strained films without dislocations. It is interesting to note that the poisson ratio of R-BFO is much larger than other perovskite materials, which have a normal value of ~ 0.3.